# State of Technology for Digital Archiving

Sandy Payette[1]
Executive Director, Fedora Commons, Inc.
Researcher in Information Science, Cornell University
November 2008

The Windsor Study Group on Digital Archiving was commissioned to recommend strategies, policies, and technologies necessary for ensuring the integrity and longevity of electronic publications. The goal of this work is to inform institutions of the challenges and opportunities faced by information stewards in fulfilling their mission of guaranteeing a permanent and authoritative scholarly record in the digital age. This white paper focuses specifically on the technological dimensions of digital archiving. It provides an analysis of the current state of technologies as well as a forecast of how digital archive systems are likely to evolve over the next decade. The thesis of this white paper is that technology does not present a barrier to long term digital archiving, but instead presents an opportunity to harness the current and future power of these technologies to create the architectural underpinnings of a comprehensive digital archiving strategy.

The state of current technologies is reflected in an array of well-known digital archives that are already serving industry, academia, and government. These are highlighted in later sections of this paper. These archives exhibit qualities that can be considered "first principles" for trusted archiving systems. They provide high-integrity storage for digital content, with mechanisms to enable fixity and bit integrity checking. They also store "essential characteristics" of content (e.g., key metadata and relationships) to ensure that critical contextual information is archived along with core content to enable effective discovery, retrieval, and management. The manner in which these essential characteristics are captured and stored varies considerably among systems, nevertheless, best practice is that this information is co-located with content, or even packaged with content.

All archive systems must offer backup and replication services to ensure that there are multiple copies of digital content. Backup, alone, does not serve as an appropriate solution for trusted digital archives. Replication of content is best practice, and it is especially important that replicas are geographically separated. To mitigate risks of technology failure, it is even better that replicated content resides in systems using different underlying technologies than the original archive system. To avoid information loss due to obsolescence of content and metadata formats, archiving systems must provide mechanisms to monitor and transform content.   Appropriate security mechanisms are essential to prevent tampering and unauthorized access to content.

---

[1] As Executive Director of Fedora Commons, I lead an organization that provides open source solutions that support digital archiving. I am also the original architect of the Fedora digital repository system mentioned in this paper. I have worked with my colleagues in the Windsor Study Group to ensure that I have provided a broad overview of the technology landscape and that I am not inadvertently biased due to my affiliation with Fedora and open source.

While most archiving solutions follow these basic principles with regard to content, there are other practices that help ensure that the systems, themselves, can live gracefully into the future.  Industry analysts at Gartner Inc. have observed, "Systems that are built to change are the only ones that last.[2]"  Gartner is leading provider of research and analysis related to the global information technology industry[3].  Practices that enable systems to evolve include an orientation toward modularity and the use of technologies and standards that enable flexibility and interoperability.  Major vendors and system integrators are embracing these practices in part to respond to customer demand, and in part to make their products and services more competitive.[4]   Indeed, Reynold Cahoon of the United States National Archives and Records Administration (NARA) observes that:

> The most basic requirement for the architecture and the technology that NARA adopts to accomplish its core mission is that the system be able to evolve over time, keeping pace with progress in IT, and responding to citizens' expectations for the best available service. Building this solution will not be easy. Based on market research conducted over the past two years, an extensive dialogue with the IT industry, and continuing collaborations with leading researchers in computer science and engineering, we have learned that key technologies enabling such a solution are available today, and that it will be possible to develop a full solution gradually over time.[5]

Beyond first principles, there are many specific and fine-grained requirements that must be fulfilled for systems to qualify as trusted digital repositories or archives.  Significant work has been done to articulate these requirements.   A well-accepted framework of archival systems is described in the Reference Model for an Open

---

[2] Charles Abrams, Roy W. Schulte, Service-Oriented Architecture Overview and Guide to SOA Research, Gartner Inc., January 2008, http://www.gartner.com/DisplayDocument?ref=g_search&id=575608&subref=simplesearch

[3] Gartner Inc., http://www.gartner.com/

[4] Patrick F. Carey and Bernard W. Gleason, Solving the Integration Issue - Service-Oriented Architecture (SOA), May 2005, http://www-03.ibm.com/industries/education/doc/content/bin/Service-OrientedArchitecture.pdf

[5] This statement was made on July 8, 2003 before the Subcommittee on Technology Policy, Information Policy, Intergovernmental Relations, and the Census of the Committee on Government Reform U.S. House of Representatives (see: http://www.archives.gov/about/speeches/07-8-03b.html)

Archival Information System (OAIS)[6]. OAIS provides a broad architectural blueprint for digital archives and it has been influential in industry, government, and academia. More recently, a number of authoritative documents have emerged that articulate audit requirements for trusted digital repositories and archives. These audit requirements address both technical and organizational aspects of archive systems. The Research Libraries Group (RLG) and NARA developed the Requirements for Trusted Digital Repositories[7], which was revised and expanded by the Center for Research Libraries (CRL) as the Trustworthy Repositories Audit and Certification (TRAC)[8]. The Digital Curation Center and Digital Preservation Europe recently released a toolkit known as Digital Repository Audit Method Based on Risk Assessment (DRAMBORA)[9]. These certification requirements and risk assessment methodologies provide both explicit and implicit requirements for the enhancing existing systems and for building the next generation of digital archiving solutions.

*Archiving Solutions: From Systems to Infrastructure*

To date, many well-designed, stand-alone systems have been built to meet the basic requirements of institutions managing specific types of digital assets. Most full-featured archiving systems provide storage, management, search, and retrieval capabilities for digital content. Often, these systems are optimized to support specific content types, where a content type is typically characterized by concrete boundaries (e.g., an article, an image, a journal) and well-known digital formats (e.g., TIFF, JPEG2000, PDF, XML, etc.).

In universities and research organizations, we increasingly see new types of scholarly and scientific artifacts that contain digital content that is media intensive (e.g., audio, video, data sets), network-based with interlinked entities, and dynamic (e.g., computational simulations). It is important to observe that new and more complex forms of scholarly output present additional challenges for digital archiving. Many of the technologies profiled in this report are mature in their abilities to handle static, well-bounded content entities but are still evolving in their abilities to handle more dynamic and complex types of content.

Some new and emerging systems are beginning to insert archival sensibilities into the front end of information lifecycles (e.g., at time of creation). Examples include deposit-oriented systems with workflows that capture preservation oriented metadata upfront, and institutional repository systems that simultaneously provide deposit, access, and preservation capabilities. Currently, a new generation of data archives is being developed to enable deposit of scientific data, while also providing tools for visualization and analysis of stored data.   All of these developments will place new demands on existing archiving systems as well as influence the direction of new archiving systems.

---

[6] OAIS, CCSDS 650.0-B-1, Blue Book, January 2002, http://public.ccsds.org/publications/archive/650x0b1.pdf
[7] OCLC Trusted Digital Repositories, http://www.oclc.org/programs/ourwork/past/trustedrep/repositories.pdf
[8] TRAC, http://www.crl.edu/content.asp?l1=13&l2=58&l3=162&l4=91
[9] DRAMBORA, http://www.repositoryaudit.eu/

*Solutions as Integrated Systems*

Does any one commercial or open source product, on its own, constitute a complete digital archiving solution?   Generally, the answer is no.   As with many large-scale systems, digital archiving systems are typically built by integrating multiple software components and core technologies, as opposed to installing a single shrink-wrapped software application.

The potential for vendor lock-in is a concern for large-scale systems.   To avoid situations where existing systems have to be abandoned or decommissioned, many institutions are motivated to build systems that can be continuously adapted and evolved by swapping out components and replacing them as needed.   As previously mentioned, building systems that can be in a continuous state of evolution is particularly important for archive systems, since the goal of the system is durability and longevity.   For maximum flexibility, institutions should strive for well-designed systems that support mass export of digital content.  Looking ahead, institutions should anticipate how digital archive systems can emerge as part of a distributed, networked infrastructure.

From a system architecture perspective, there are several well-established system integration approaches that used in building digital archiving systems:  tightly coupled component-based systems, service oriented architectures (SOA) serving an institution or enterprise, and loosely coupled distributed architectures (distributed web services, GRID).   Interestingly, systems can exhibit qualities of more than one of these approaches.  For example, enterprise SOA can include connectors to tightly coupled systems and may also have distributed components.   It is also notable that distributed architectures can show characteristics of emerging infrastructure, in particular the interconnection of components, services, and sub-systems into a super structure that is robust, durable, and evolvable over time.

Currently, digital archiving is supported by a diversity of commercial and open source technologies, provided by both for-profit and not-for-profit institutions.  This is a healthy development that helps mitigate the risk of monopolies that could negatively affect universities in fulfilling their mission of preserving the scholarly record.   What follows is a brief review of commercial and open source offerings that support digital archiving.

*Commercial Products*

Commercial enterprise content management systems are often marketed as digital archiving solutions. Examples include Documentum[10], ContentDM[11], and IBM Enterprise Content Management[12] as well as product suites by many other major vendors.[13] We often see these products integrated with other technologies in the development of full-featured enterprise systems. In addition to content management products, there are also commercial offerings that focus exclusively on the storage aspects of digital archiving and thus require enterprise IT teams or system integrators to build the rest of the archive system upon these storage-centric products. Examples targeted at the digital archiving market include Sun Honeycomb[14], Sun Open Storage[15], EMC Centera[16], and Symantec Enterprise Vault[17]. Many commercial products have proven their capabilities by serving the business archives market, primarily focused on documents and email, and also serving the commercial content market with a focus on multimedia content, especially audio and video.

Some commercial systems are optimized to deal with specific types of content. In these cases, the system may not be easily adapted to accommodate the requirements of new or more specialized forms of digital content. As previously mentioned, new expressions of digital scholarship may present unanticipated challenges to existing content management solutions. Also, in the academic sector, there are concerns as to whether commercial products can be adapted to support community-specific standards. For example, there are several content packaging and encoding formats that are favored by libraries for ingest and export of digital content (METS[18], MPEG21[19], IMS[20], NLM[21]). Similarly, libraries are strong advocates of several community-specific metadata standards (e.g., Dublin Core[22],

---

[10] Documentum, http://www.documentum.com/
[11] ContentDM, http://www.contentdm.com/
[12] IBM ECM, http://www-01.IBMibm.com/software/data/content-management/
[13] See article by Karen M. Shegda, et al. entitled the Magic Quadrant for Enterprise Content Management, Gartner, September 2008, http://mediaproducts.gartner.com/reprints/microsoft/vol6/article3/article3.html
[14] Sun Honeycomb, http://www.sun.com/storage/disk_systems/enterprise/5800/
[15] Sun Open Storage, http://www.sun.com/storage/openstorage/
[16] EMC Centera, http://www.emc.com/products/family/emc-centera-family.htm
[17] Semantic Enterprise Vault, http://www.symantec.com/business/enterprise-vault
[18] Metadata Encoding and Transmission Standard (METS), http://www.loc.gov/standards/mets/
[19] MPEG21, http://www.chiariglione.org/mpeg/working_documents.htm#MPEG-21
[20] IMS Content Packaging, http://www.imsglobal.org/content/packaging/
[21] NLM Journal Archiving and Interchange Tag Suite, http://dtd.nlm.nih.gov/
[22] Dublin Core Metadata Initiative, http://dublincore.org/

MODS[23] for descriptive metadata, and PREMIS[24] for preservation metadata) and will want these to be accommodated by digital archiving systems.

*Open Source Software*

Open source software solutions can provide an alternative to commercial products or they can coexist with commercial products as part of an integrated system. The primary benefits of open source include openness, transparency, and community-based control over the future direction of the software.   A key aspect of open source is that program code and data structures are visible for inspection and open for modification.   This enables an institution to make changes to the software as needed.  It also permits communities to collectively evolve the software to meet new or specialized requirements.   It is beneficial to use open technologies for archival system components that are most closely tied to the storage and management of digital content.   In these key areas, open source can help mitigate the risk of technology monopolies in areas that most closely control digital content.

A number of open source solutions have emerged that address some or all of the basic requirements of digital archiving including Fedora[25], DSpace[26], EPrints[27], aDORe[28], LOCKSS[29], iRODS[30].  While each system has its own strengths and weaknesses, all of them offer a more open approach than do commercial systems.  Not-for-profit corporations have been established to take on responsibilities of sustaining and evolving several of these solutions, notably Fedora Commons, the DSpace Foundation, and iRODS (currently in formation).   From a sustainability standpoint, it is important to observe that community support (either in-kind or financial) is the key to the long-term viability of any open source offering.

Open source digital repository systems have been used for many purposes, including digital archiving.  Generally, digital repository systems serve as core components within an archive system and they are typically integrated with other commercial or open-source components.   Fedora is a digital repository system that is notable for its modularity, flexibility, and scalability.  Fedora was featured in a prototype of the Electronic Records Archive for the United States National Archives and Records Administration (NARA)[31]. DSpace offers a turnkey application for institutional repositories and has significant uptake in university libraries for deposit of scholarly materials by faculty. DSpace has provided libraries an easy and affordable entry point for creating small to medium sized preservation repositories. EPrints began as a system to provide open access to pre-prints of scholarly publications and has also evolved to be a more general-purpose repository system for research outputs.

---

[23] Metadata Object Description Scheme (MODS), http://www.loc.gov/standards/mods/
[24] PREservation Metadata: Implementation Strategies (PREMIS), http://www.oclc.org/research/projects/pmwg/
[25] Fedora Commons, http://www.fedora-commons.org
[26] DSpace, http://www.dspace.org/
[27] EPrints, http://www.eprints.org/
[28] aDORe Archive, http://african.lanl.gov/aDORe/projects/adoreArchive/
[29] Lots of Copies Keeps Stuff Safe (LOCKSS), http://www.lockss.org
[30] Integrated Rule-Oriented Data System (iRODS), http://www.irods.org

Other open source solutions are targeted more precisely at digital archiving. LOCKSS (Lots of Copies Keeps Stuff Safe) offers a peer-to-peer replication infrastructure for preserving digital content. Many major universities have used LOCKSS as a solution for electronic journal archiving. The aDORe archive system was developed to provide a highly scalable, standards-based solution for archiving electronic publications for the Los Alamos National Laboratory (LANL) and has since been made available to others as open source. The iRODS system is notable for its rule-based approach to digital preservation. User communities can assert content management policies and the iRODS rule engine will enforce them. iRODS is a successor to the Storage Resource Broker (SRB) which was the basis for several NARA test beds, and also used by digital library and scientific data grid communities.

*Emerging Infrastructure*

A historical look at the emergence of infrastructure - electric grids, railways, and the Internet – shows that a key developmental stage is when formerly incompatible or stand-alone systems are interconnected via adapters and gateways.[32] We see this happening with some aspects of digital archiving systems. First, there is convergence in the area of core storage technologies where standards exist for connecting heterogeneous storage devices to form local or distributed storage fabrics (i.e., SANs, GRID, and Cloud discussed later). In the digital repository domain, the major open source repository communities are in active discussions on how heterogeneous repository systems can evolve to become components of a broader networked infrastructure. For example, Fedora, DSpace, and EPrints are experimenting with protocols and formats to enable exchange of content among them.

The distributed peer-to-peer nature of LOCKKS already demonstrates a community-based distributed infrastructure for replication of content. The GRID[33] initiatives

---

[31] The Fedora-based prototype was developed by Harris Corporation, one of two finalists for the NARA bid to develop the Electronic Records Archive. The contract was ultimately awarded to the other finalist, Lockheed Martin.

[32] See report by Paul Andrews, et al. entitled "Understanding Infrastructure: Dynamics, Tensions, and Design", January 2007, http://hdl.handle.net/2027.42/49353

[33] What is the Grid?, www-fp.mcs.anl.gov/~foster/Articles/WhatIsTheGrid.pdf

demonstrate how globally distributed scientific communities have cooperated to develop a high quality-of-service (QoS) infrastructure for computation and storage[34], with significant successes, but also some criticism of excessive complexity.  As distributed infrastructures emerge, we also see efforts to create global authorities for shared information.   For example, there are two major projects focused on digital format registries, most notably the Global Digital Format Registry (GDFR)[35] and the UK National Archive's PRONOM registry[36]

All of this suggests that digital archiving technologies are on a path towards emergent infrastructure.   This trend offers the prospect of a future where multiple institutions collaborate or form virtual organizations to leverage common infrastructure for housing large-scale, distributed digital archives.  The benefits are that institutions can share technologies and more easily achieve economies of scale.  This can become a reality with appropriate commitment and financial investment by institutions with archival responsibilities as well as through national and government funding.  The emergence of national and international infrastructure programs is an important step in enabling this to happen.  In the United States, the National Science Foundation's Office of Cyberinfrastructure[37] has commissioned significant reports and has begun funding new and innovative work.   The Library of Congress has worked with the NSF to create the National Digital Information Infrastructure and Preservation Program (NDIIPP)[38] program.  In the UK the Joint Information Systems Committee (JISC) has established a set of inter-related programs that include the Repositories and Preservation Programme[39], the Digital Preservation and Records Management Programme[40], and the e-Infrastructure Programme[41]

### *Maturity:  Where Are the Underlying Technologies Now?*

Technologies supporting digital archiving have matured significantly over the past decades due to significant investment by both government and industries motivated to ensure long-term access and preservation of digital assets  (e.g., government records, media assets including digital video and images, scientific data).   As previously mentioned, any worthy digital archiving system will require a core set of underlying technologies that provide storage, backup and recovery, replication, format migration, search and retrieval.  The state of these core technologies is described briefly below.

---

[34] Grid Storage, http://www.gridpp.ac.uk/wiki/Grid_Storage
[35] Global Digital Format Registry (GDFR), http://www.gdfr.info/
[36] PRONOM, http://www.nationalarchives.gov.uk/pronom/
[37] NSF Office of Cyberinfrastructure, http://www.nsf.gov/dir/index.jsp?org=OCI
[38] NDIIPP, http://www.digitalpreservation.gov/library/
[39] JISC Repositories and Preservation Programme, http://www.jisc.ac.uk/whatwedo/programmes/reppres
[40] JISC Digital Preservation and Records Management Programme, http://www.jisc.ac.uk/whatwedo/programmes/preservation.aspx
[41] JISC e-Infrastructure, http://www.jisc.ac.uk/whatwedo/programmes/einfrastructure

*Storage*

Large scale, reliable storage is essential for digital archive systems.  Fortunately, core storage technologies are quite mature and reasonably standardized.  Modern deployments typically consist of Storage Area Networks (SAN) connecting servers to storage arrays using either Fibre Channel (FC) or Internet Small Computer System Interface (iSCSI) networks.  Fibre Channel requires dedicated cabling and is the dominant standard in enterprise storage for connecting SANs.  In contrast, iSCSI can be run over existing Internet infrastructure using Internet protocols.

Direct attached storage is typically not used for enterprise or archival storage.  Instead, it is more common to see large storage systems hosted in data centers and shared by multiple organizational units or even outside users.  In this scenario, storage can be partitioned so that it appears that it is dedicated to particular organizational unit or project, thus reducing concerns of privacy while still obtaining economies of scale and organizational flexibility.

Concerns about vendor lock-in are mitigated by the standardization of network connection protocols and input/output interfaces for storage systems (e.g., ANSI T11[42], T10[43], IETF RFC 3720[44]).  The good news for digital archiving systems is that, typically, storage components from different vendors can be used simultaneously, or swapped in and out over time.  With effective planning organizations can ensure smooth evolutionary upgrades or migrations of storage components and subsystems.

A recent trend is the emergence of "cloud" storage providers that offer storage as a Web-based service backed by large data centers.  Well-known examples include Amazon S3[45], Google Data[46], and Microsoft Azure[47].  There are also new players such as Nirvanix and Atemp[48] that have partnered to offer cloud-based file archiving.  Other major companies are developing cloud strategies including Hewlett Packard and Yahoo[49].  While this trend may be of interest to organizations building digital archives, it should be noted that, currently, cloud vendors offer simplified, less robust interfaces and lower Service Level Agreements (SLA) than are appropriate for long-term digital archives. Also cloud vendors do not provide high-level security infrastructures at this time. While cloud storage providers are likely to be players in the archiving space, there is still work to be done to address issues such as data loss, integrity guarantees, and security.

---

[42] ANSI T11 standard, http://www.t11.org/index.html
[43] SCSI and SATA standards, http://www.t10.org/
[44] IETF RFC 3720, http://tools.ietf.org/html/rfc3720
[45] Amazon Simple Storage Service (S3), http://aws.amazon.com/s3/
[46] Google Data APIs, http://code.google.com/apis/gdata/overview.html
[47] Microsoft Azure, http://www.microsoft.com/azure/default.mspx
[48] Nirvanix is a startup offering cloud-based file archiving, http://searchstorage.techtarget.com/news/article/0,289142,sid5_gci1319281,00.html
[49] See announcement of cloud testbed by HP, Intel, Yahoo, www.hp.com/hpinfo/newsroom/press_kits/2008/cloudresearch/index.html

*Backup and recovery*

Traditional backup and recovery methods, usually tape, may be used for smaller content systems or archive implementations (less than a petabyte). The common methodology consists of periodic complete backups followed by a series of incremental backups of changed material. Sun offers the Storage and Archive Manager (SAM)[50], which is tightly integrated with the Quick File System (QFS)[51]. Known together as SAM/QFS, this is a solid hierarchical storage management approach for archival storage where content is stored on tape backup and retrieved to disk when access is required. It also has the advantage of being open source. Other mature backup solutions are provided by EMC[52], Veritas[53], and IBM[54].

*Replication*

Large storage systems (petabyte and larger) cannot easily use traditional backup techniques. Generally, a replication strategy is used as a substitute, which has the benefit of enabling both high availability and content preservation strategies. In general, it is best if the replicas are geographically distributed. Two approaches may be employed: file replication or entity replication. File replication is the most mature approach. The files (byte streams) on a storage subsystem are copied to a remote location. Example products include SAM/QFS, Veritas Volume Replicator[55], and Brocade File Management[56].

---

[50] Sun Storage and Archive Manager (SAM), http://www.sun.com/storagetek/management_software/data_management/sam/
[51] Sun Quick File System (QFS), http://www.sun.com/storagetek/management_software/data_management/qfs/
[52] EMC Backup, Recovery and Archiving, http://www.emc.com/solutions/business-need/backup-recovery-archiving/index.htm
[53] Veritas NetBackup, http://www.symantec.com/business/netbackup
[54] IBM Tivoli, http://www-01.ibm.com/software/tivoli/products/storage-mgr
[55] Veritas Volume Replicator, http://www.symantec.com/business/volume-replicator
[56] Brocade File Management, http://www.brocade.com/products-solutions/products/file-management/index.page

Entity replication entails copying sets of related byte streams (i.e., "entities"). Key information is also copied such as metadata and information about entity organization that is essential for recreating access to entities. Entity replication is a large emerging market that is heavily driven by business use cases. Examples are IBM Records Manager[57], Symantec Enterprise Vault[58] and EMC Centera[59]. It should be noted that most current commercial offerings for entity replication suffer from interoperability problems and a limited ability to replicate content between vendors. A new standard known as the eXtensible Access Method (XAM) may lead to more interoperability among different vendors offering products in this space[60].

On the open source front, LOCKSS is directly addressing replication of content. Looking ahead, Fedora Commons and the DSpace Foundation are collaborating on the conceptualization and design of an open source storage overlay service that can be used to replicate content across heterogeneous storage providers, both enterprise and cloud-based.

*Content migration*

The ability to migrate digital content to new formats is an essential capability of a digital archive system. The basic process of migration requires a means of evaluating current formats and executing processes to transform content items. Identification of the current format of a content item can be accomplished in different ways, including periodically reflecting on key attributes of content items, querying a registry of content items, or using a service that can figure out formats by analyzing content byte streams. Once content formats are identified and verified, there must be a way to consult rules or policies that reveal whether a content format is endangered, obsolete, or undesirable for other reasons. Migration decisions can be made in automated, semi-automated, and non-automated ways. Ultimately, an archive system must be able to initiate, execute, and validate transformations of content items from original formats to new formats.

There is no single product that supports migration for *all* common content formats. A mature commercial product is Oracle's Outside In[61], which supports 400 formats with converters and/or viewers. A commonly used tool in open source is Harvard's

---

[57] IBM Records Manager, http://www-01.ibm.com/software/data/cm/cmgr/rm/
[58] Symantec Enterprise Vault, http://www.symantec.com/business/theme.jsp?themeid=datacenter
[59] EMC Centera, http://www.emc.com/products/family/emc-centera-family.htm
[60] eXtensible Access Method (XAM), http://www.snia.org/forums/xam/
[61] Outside In, http://www.oracle.com/technologies/embedded/outside-in.html

JHOVE[62]. There are other smaller conversion products available each handling a relatively small number of formats. A large number of formats can be converted to PDF and the archival format PDF-A. Adobe supplies these products in Acrobat[63] but converters are available from many other sources. Transformation of XML data is very well supported and standardized with a plethora of vendors providing technologies for XML query and transformation (e.g., XSLT, XPath, XQuery, and XSL-FO). Converters between most versions of Microsoft products are available from Microsoft or other vendors.

Extract Transform and Load (ETL) tools support transformations for structured data especially relational databases (RDBMS). In open source Enhydra Octopus[64] provides a framework for ETL. Oracle Data Integrator and SAS provide powerful commercial ETL tools.

*Archive Search*

An essential part of accessing and retrieving material stored in digital archives is the ability to find items of interest. Search is generally identified as a technology that is able to extract and index features from digital content, or index metadata about content. Examples include commercial search engines such as FAST[65], Autonomy[66], and IBM Omnifind[67]. Open source products include Lucene[68], Solr[69], and Zebra[70].

Search is a mature technology and continued advances in search engines are anticipated. Search engines are easy to fit into digital archiving architectures since they are able to index material from practically any source. Search engines are relatively low risk component in digital archives, since they can be replaced and indexes can be rebuilt using processes that run over the original archive to re-index content. One area that can sometimes present a challenge is the application of access controls and security filters to searches when sensitive or proprietary information finds its way into search indexes.

A clear distinction should be drawn between enterprise search and public search engines. Enterprise search is distinguished in that it provides customized indexing and security under the control of the organization that is responsible for the system. Public search (e.g., Google) is largely out of the organization's control but may be desirable for exposure of materials in open public archives.

---

[62] JHOVE, (http://hul.harvard.edu/jhove).
[63] Adobe Acrobat, http://www.adobe.com/products/acrobat/
[64] Enhydra Octopus, http://www.enhydra.org/tech/octopus/index.html
[65] FAST Search, http://www.fastsearch.com/
[66] Autonomy Search, http://www.autonomy.com
[67] Omnifind, http://www-01.ibm.com/software/data/enterprise-search/omnifind-enterprise/
[68] Lucene, http://lucene.apache.org/java/docs/
[69] Solr, http://lucene.apache.org/solr/
[70] Zebra, http://www.indexdata.dk/zebra/

*Technologies in Action: Notable Digital Archives*

A number of well-known archive solutions exist that specifically target electronic journals and scholarly literature. Portico[71] is a non-profit organization whose mission is to preserve scholarly literature published in electronic form. Portico hosts an archive system whose development was influenced by the OAIS framework and handles file storage, backup, replication, and migration. Portico developed an integrated system that uses standards such as METS and PREMIS. It also integrates a well-known commercial content management product (Documentum). JSTOR[72] is another non-profit organization whose mission includes serving the scholarly community by providing a trusted digital archive of high quality scholarly journals. JSTOR provides access to journals, but it also preserves electronic editions of journals using Portico. The National Library of Netherlands developed an archiving solution known as e-Depot[73] and is currently offering it as a service to publishers worldwide. Another approach to shared archiving is seen with the formation of CLOCKSS[74], a not-for-profit partnership of libraries and publishers to run a cooperative archive using the previously mentioned LOCKSS technologies. CLOCKKS recently completed a prototype phase of the cooperative archive, and plans to go live in January 2009.

The Internet Archive[75] is a non-profit organization established to preserve web sites by taking regular snapshots of the web. It has expanded to include other types of digital collections including images, books, and videos. Internet Archive also houses collections of digitized texts from research libraries that are members of the Open Content Alliance[76]. Recently, NASA and Internet Archive announced a partnership to provide public access and long-term preservation of NASA's vast collection of photographs, historic film and video.[77]

Archiving is a concern of government agencies and there is some public information about existing and emerging archiving systems in this domain. One notable archive is the Comprehensive Large Array-data Stewardship System (CLASS)[78] that is run by the National Oceanic and Atmospheric Administration (NOAA). The CLASS archive is a custom built integrated system that archives weather and atmospheric data. Currently, the National Archives and Records Administration is building its new Electronic Records Archive (ERA). The system is being developed by Lockheed Martin (general contractor) and will likely integrate commercial products, custom developed components, and some open source components.

---

[71] Portico, http://www.portico.org/
[72] JSTOR, http://www.jstor.org/
[73] e-Depot, http://www.kb.nl/dnp/e-depot/e-depot-en.html
[74] CLOCKKS, http://www.clockss.org/clockss/Home
[75] Internet Archive, http://www.archive.org/
[76] Open Content Alliance, http://www.opencontentalliance.org/
[77] NASA and Internet Archive press release, http://www.archive.org/iathreads/post-view.php?id=201294
[78] NOAA CLASS, www.class.noaa.gov/

Many university libraries and IT organizations have taken a leadership position in digital preservation and digital archiving. Universities that are running their own digital archive systems include University of Michigan, Stanford University, California Digital Library, John Hopkins University, University of Virginia, Bibliotheque de France (BNF), National Library of Australia, Oxford University, and others. The OAIS framework has influenced many of these universities. Cornell University Library and the University of Michigan are especially known for their early work in digitization and digital preservation.

### *Conclusion: Opportunities for Higher Education*

Universities have an opportunity to achieve economies of scale by collectively engaging in the mission of archiving the scholarly record. At this time, universities can work together to determine how best to position themselves in the dynamic environment of technologies to maximize benefits and mitigate risk. It is important for leaders in higher education to understand the dynamics and tensions of emerging infrastructure and to strategize on how to leverage it effectively, not be disrupted by it. In the recent EDUCAUSE e-book entitled *The Tower and the Cloud*[79], Richard Katz and others discuss the implications of the networked information economy and how it will impact higher education institutions. Reflecting on the possibilities of how institutions can position themselves, Katz observes,

> This evolution presents enormous opportunity and risk. We can leave our institution's response to these new capabilities to chance, we can confine our planning to the IT organization, or we can engage our leadership in a discussion of near-term strategy. The strategic discussion about the evolution of virtualization, services orientation and delivery, open resources, web standards, and cloud computing is in fact a conversation about what constitutes the institution as an enterprise and about how the institution wishes to manifest its institutional presence in cyberspace.[80]

There are many options for universities in establishing digital archive systems for electronic publications, datasets, and other forms of scholarly materials in digital form.   These run the range from building or buying archiving systems to support a single university or unit within, to collaboratively archiving content through cooperative partnerships with other universities and institutions (e.g., LOCKSS), to outsourcing archiving to external providers (e.g., Portico), and, ultimately, to developing new strategies and building archives that leverages shared network infrastructure in new ways.

---

[79] Richard Katz is editor of a new book *The Tower and the Cloud* released under the Creative Common license at EDUCAUSE in October 2008, http://www.educause.edu/thetowerandthecloud/133998

[80] Richard Katz, "The Gathering Cloud: Is This the End of the Middle", chapter in *The Tower and the Cloud*, p. 22.

Whichever direction universities take in the short term, they should be committed to ensuring that digital archiving systems exhibit fitness of purpose as described in the first principles and requirements for trusted digital archives referred to in this paper.   To review, the basics include providing high integrity storage for digital content, including the ability to capture and store all essential characteristics of the content.  It is also essential to ensure that multiple copies of content exist and that copies are stored in different locations.   It is best that copies are geographically dispersed, and it is even better if copies are stored on systems built with different underlying technologies.

Well-designed digital archive systems have qualities that enable them to evolve, meaning that over time technology components can be swapped in and out as needed.  When this is achieved, the archive system takes on an organic quality in which the system can endure as a trusted archive for the digital content that the system is commissioned to protect.   Also of key importance for an archive system is the ability to mass export digital content, along with all key metadata and relationships that are essential to making content usable.   Evolvable systems are the ideal for digital archiving systems and exit strategies are essential.

The term "virtualization" refers to the abstraction of computing resources resulting in storage and applications giving the appearance of being virtually in one place, but actually being distributed over an array of devices and servers connected via networks.  These networks can exist within an institution or can be geographically dispersed.   Virtualization is the phenomenon that's behind cloud computing, and also found within enterprises and throughout commercial data centers.   Storage fabrics (e.g., SANs, GRID storage) such as those discussed early in this paper exemplify virtualization and make it possible to connect heterogeneous storage devices supported by different hardware and software platforms.   From a computing perspective, virtualization makes it possible to use the computing power of many servers regardless of the physical location.

Virtualization technologies are significant for digital archiving systems, most notably for large-scale digital archives.  Virtualization presents the possibility of organizations or universities joining together to be able to share networked devices (e.g., storage, software services, compute power) with to better manage costs, security, power consumption, and policy.   As suggested by Katz, in the future, local control in higher education information technologies may not be at the enterprise but instead it may be distributed across multiple institutions, and will likely involve dependencies on emergent players such as providers of cloud computing resources.

Technology offers great prospects for the future of digital archive systems; at the same time, technology alone cannot be the sum total of a trusted digital archiving solution.   The quest to archive digital content must always be achieved with a multi-dimensional strategy that addresses not just technologies but the organizational, economic, and social realities of universities.   Long-term success of digital archiving can be ensured only when ongoing *audit* processes that measure the state of digital archive systems and the state of the organizations that support the archives are in place.  This is another opportunity for multi-institutional collaboration - the design and development of robust and authoritative audit mechanisms.   By working together on both the technology and audit dimensions of digital archiving, universities have an opportunity to make significant progress solutions for preserving the scholarly record for future generations.

***Acknowledgements***

I would like to thank members of the Windsor Study Group for their valuable suggestions and feedback, especially James Hilton, Don Waters, and Paul Courant.  Also, I would like to thank Dan Davis of Cornell University for his assistance on the research for this paper, and for his perspectives on large-scale digital archiving systems informed by his work as chief architect for the Harris Corporation prototype of the NARA Electronic Records Archive.